\documentclass[aps,pre,twocolumn]{revtex4}
\usepackage{amssymb}
\usepackage{graphicx}
\usepackage{colordvi}

\begin{document}
\title{Collective dynamics of Fermi-Bose mixtures with an oscillating scattering length}
\author{F. Kh. Abdullaev$^{1,2}$, M. \"{O}gren$^{3,4}$, M. P. S\o rensen$^{5}$}

\date{\today{}}

\address{$^{1}$Physical-Technical Institute, Uzbek Academy of Sciences, 2-b G. Mavlyanov Street, 100084 Tashkent, Uzbekistan\\
$^{2}$CCNH, Universidade Federal do ABC, 09210-170, Santo Andr\'e, Brazil\\
$^{3}$School of Science and Technology, \"{O}rebro University, 701
82 \"{O}rebro, Sweden\\
$^{4}$Nano Science Center, Department of Chemistry, University of
Copenhagen, Universitetsparken 5, 2100 K{\o}benhavn {\O}, Denmark\\
$^{5}$Department of Applied Mathematics and Computer Science, Technical University of Denmark, 2800 Kongens Lyngby, Denmark}

\begin{abstract}
Collective oscillations of superfluid mixtures of ultra cold fermionic and bosonic atoms are investigated while varying the fermion-boson scattering length.
We study the dynamics with respect to excited center of mass modes and breathing modes in the mixture.
Parametric resonances are also analyzed when the scattering length varies periodically in time, by comparing
partial differential equation (PDE) models and ordinary differential equation (ODE) models for the dynamics.
An application to the recent experiment with fermionic $^{6}$Li and bosonic $^{7}$Li atoms, which approximately have the same masses, is discussed.
\end{abstract}
\maketitle

\section{Introduction}

Fermi-Bose mixtures have attracted much attention lately, starting with the work~\cite{Schreck2001}, where the first observation of the mixture of a Bose-Einstein condensate in a Fermi sea has been performed.
Among the different types of Fermi-Bose mixtures, particular interest is devoted to a mixture where both components are superfluids. One important example is the $^{6}$Li-$^{7}$Li mixture~\cite{Abeelen,Kempen}.
Recently, experimental production of a mixture of fermionic and bosonic atoms, where both components are superfluids, has been reported for $^{6}$Li-$^{7}$Li mixture~\cite{Ferrier, Delehaye}.
Other interesting systems are the $^{40}$K-$^{41}$K mixture with a tunable interaction between species~\cite{Wang,Falke,CWu} and the mixture of $^{133}$Cs and $^{6}$Li with broad interspecies Feshbach resonances~\cite{Repp,Tung}.
Recently, the two-species superfluid $^6$Li-$^{41}$K and $^6$Li-$^{176}$Yb has been experimentally realized~\cite{Yao,Roy}.
Among the new effects predicted for mixtures of Bose-Fermi superfluids we mention the existence of vortices in a rotating quasi-two-dimensional Fermi-Bose mixtures~\cite{WenPRA2014}, Faraday waves~\cite{Abdullaev2013}, an prediction of the existence of super-counter-fluid phase~\cite{Kuklov}, the existence of dark-bright solitons~\cite{MitraJLTP2018}, and the multiple periodic domain formation~\cite{TylutkiNJP2016}.
Collective oscillations of the Fermi-Bose mixtures for different settings has earlier been analysed in~\cite{Ferliano,Banerjee,Nascimbene,Wen}.
Dipole modes in the BCS-BEC crossover regime has been analytically considered in~\cite{Wen}.
In the recent work~\cite{Wu} the dipole oscillations of the Fermi-Bose-mixture with a large mass imbalance have been experimentally investigated.
It was observed that when the inter-species interaction strength is varied, the dipole oscillations frequency shifts and the $^{41}$K and $^6$Li components display a resonance like behavior in the upward and downward (radial and axial) directions.

The aim of the present paper is to investigate the collective dynamics of the Fermi-Bose mixture when the scattering length is varied periodically in time.
We start in section II with a description of the model.
To analyze the collective oscillations we then introduce a variational approach.
In section III we investigate the dynamics for the center of mass and for breathing modes.
There are different possible resonant regimes depending on what kind of scattering length that are varied.
In particular the variation of the fermion-boson scattering length $a_{s,fb}$ will lead to parametric resonance in the oscillations of the relative distance between the centers of mass of both superfluids.
Also we study the resonance dynamics of breathing modes of both superfluids.
The numerical simulations are discussed in section IV.
Finally we conclude in section V that the resonances should be clear experimental signals to search for.
The details of the variational approach are described in the Appendix.

\section{The model} \label{sec_The_model}

The time-dependent model is described by the system of partial differential equations (PDEs) \cite{Adhikari,Abdullaev2013}
\begin{eqnarray}
i\psi_{b,t} &=& -\psi_{b,xx} + \alpha_b x^2\psi_b + g_b |\psi_b|^2\psi_b + g_{fb}|\psi_f|^2\psi_b \; , \nonumber \\
i\psi_{f,t} &=& -\psi_{f,xx} + \alpha_f x^2\psi_f + \kappa \pi^2 |\psi_f|^4\psi_f + g_{fb}|\psi_b|^2\psi_f \; , \nonumber \\ \label{sys1}
\end{eqnarray}

\noindent where the parameters $g_b, g_{fb}$ are proportional to atomic scattering lengths $a_{s,b}, a_{s,fb}$, respectively, while $\alpha_b, \alpha_f$ determines the harmonic trapping.
Furthermore, $g_{b}= 2\hbar a_{s,b} \omega_{\perp}$ is called the one-dimensional coefficient of mean-field
nonlinearity for bosons, where $a_{s,b}$ is the scattering
length and $\omega_{\perp}$ is the perpendicular frequency of the
trap. Similarly, $g_{fb}$ is called the interspecies interaction coefficient~\cite{Adhikari}.
We are interested in weak Bose-Bose interactions (we consider small positive $g_b$) and attractive Fermi-Fermi interactions such that the superfluid Fermi-Bose system is described by the nonlinear Schr\"odinger-like equation~(\ref{sys1})~\cite{Heiselberg,Bulgac,Manini,Astr,Adhikari2}.
In the BCS weak attractive coupling limit
the  fermionic subsystem coefficient is $\kappa =1/4$, while in the molecular unitarity limit
it is $\kappa=1/16$~\cite{Adhikari}.
Finally, for the bosonic Tonks-Girardeau limit \cite{1DTonks}
with the components $\psi_b$ ($\psi_{f}$) being a weakly (strongly) repulsive bosonic species
we have $\kappa =1$.
The system~(\ref{sys1}) is written in dimensionless form using the variables
\begin{eqnarray}
&& l_{\perp} = \sqrt{\frac{\hbar}{m_b \omega_{\perp}}} \: , \; \psi_{b,f} = \sqrt{l_{\perp}} \Psi_{b,f} \: , \; t=\tau \omega_{\perp} \; , \nonumber \\
&& x = \frac{X}{l_{\perp}} \: , \; g_{n} = \frac{2m_b l_{\perp}}{\hbar^2}G_{n} \: \nonumber
\end{eqnarray}

\noindent where $G_n$ ($n=b,fb$) is the coefficient of the mean-field nonlinearity.
We also implicitly assume that $m_b = m_f$ in Eq.~(\ref{sys1}).
As mentioned, such a condition is approximately realized in the $ ^{6}$Li-$^7$Li and $^{40}$K-$^{41}$K mixtures.

Below we will also consider the situation of a sinusoidal time dependence of the scattering length $a_{s,fb}$.
This can be achieved using the Feshbach resonance technics \cite{Kagan, Kagan2, Inouye, Collective}.
According to these schemes, we can manipulate the scattering lengths by varying an external magnetic field in time near the resonant value~\cite{Ferliano06}.
For the Fermi-Bose mean-field nonlinearity in the model~(\ref{sys1}) it means that the interaction parameter will be varied in time as
\begin{equation} \label{TDgbANDg12}
g_{fb}(t) = g_{fb}^{(0)}[1 + c_{fb} \sin(\Omega_{fb} t)] \; .
\end{equation}

To analyse collective oscillations of the mixture with the scattering lengths varying in time, we will employ the variational approach (VA).
According to this method we first calculate the averaged Lagrangian
\begin{equation}\label{avlag}
\bar{L} = \int_{-\infty}^{\infty}L(x,t)dx \; ,
\end{equation}

\noindent where $L(x,t)$ is the Lagrangian for the system (\ref{sys1})
\begin{eqnarray}\label{lag}
L(x,t) &=& \sum_{n=b,f} \left[ \frac{i}{2}(\psi_n\psi_{n,t}^{\ast}-c.c.) +|\psi_{n,x}|^2  +\alpha_n x^2 |\psi_n|^2 \right] \nonumber\\
&& + \frac{g_b}{2}|\psi_b|^4 + \frac{\kappa\pi^2}{3}|\psi_f|^6 + g_{fb}|\psi_b|^2|\psi_f|^2 \; .
\end{eqnarray}

\noindent For the Bose and Fermi wave functions we employ the Gaussian ansatz
\begin{equation}\label{VA}
\psi_n = A_n \exp \left( -\frac{(x-\zeta_n)^2}{2a_n^2} \right)
e^{i(b_n(x-\zeta_n)^2 + k_n(x-\zeta_n)+\phi_n)} \; , \nonumber \\
\end{equation}

\noindent for $n=b,f$ and where the parameters $A_n$, $a_n$, $b_n$, $k_n$, $\zeta_n$, $\phi_n$ are all
real functions of time $t$.
Substituting the ansatz~(\ref{VA}) into Eq.~(\ref{lag}) and performing the average in Eq.~(\ref{avlag}), we obtain the following averaged Lagrangian
\begin{eqnarray}\label{avlag1}
\bar{L}(t)&=&\sum_{n=b,f} \Bigl\{ \sqrt{\pi}A_n^2 \left( \frac{1}{2a_n} + 2b_n^2 a_n^3 + a_n k_n^2 \right.\nonumber\\
&+& \left. \frac{1}{2}a_n^3 b_{n,t} - a_n k_n\zeta_{n,t} + a_n \phi_{n,t} \right) + \frac{\sqrt{\pi}\alpha_n}{2}A_n^2 a_n^3   \nonumber\\
&+& \sqrt{\pi}\alpha_n A_n^2 a_n \zeta_n^2 \Bigl\} + \frac{\sqrt{\pi}g_b}{2\sqrt{2}}A_b^4 a_b  + \frac{\pi^{5/2}\kappa A_f^6 a_f}{3\sqrt{3}} \nonumber\\
&+& \frac{\sqrt{\pi}g_{fb}a_b a_f A_b^2 A_f^2}{\sqrt{a_b^2 + a_f^2}}e^{-\frac{(\zeta_b - \zeta_f)^2}{a_b^2 + a_f^2}} \; .
\end{eqnarray}

\noindent The numbers of atoms corresponding to the Gaussian wavefunctions are equal to
\begin{equation}
N_n = \int_{-\infty}^{\infty} |\psi_n|^2 dx = \sqrt{\pi}A_n^2 a_n \; ,  \; \; n = b,f \; .
\end{equation}

\noindent Variation of the Lagrangian in (\ref{avlag1}) with respect to $\phi_b$ and $\phi_f$
shows that $N_b$ and $N_f$ are constants. The dynamical equations for the parameters, or collective coordinates,
$\xi_i \in$ $\{ \zeta_n$, $k_n$, $b_n$, $A_n$, $a_n \} $ with $n=b,f$ can be derived from the
Euler-Lagrange equations for the averaged Lagrangian~(\ref{avlag1}), leading to the system of ordinary differential
equations (ODEs)

\begin{equation}
\frac{\partial\bar{L}}{\partial\xi_i} = \frac{d}{dt}\frac{\partial\bar{L}}{\partial \xi_{i,t}} \; . \label{Euler-Lagrange-Equation}
\end{equation}

\noindent Variation with respect to $\phi_n$ has already been taken care off
leading to a constant $N_n$.

\section{Collective dynamics}

\subsection{Centers of mass oscillations modes }

For the center of mass coordinates ($\zeta_b,\: \zeta_f$), we obtain from (\ref{Euler-Lagrange-Equation}) the coupled differential equations (see the Appendix for details)
\begin{eqnarray}
\zeta_{b,tt} &=& -4\alpha_b \zeta_b + \frac{4g_{fb}N_f}{\sqrt{\pi}(a_b^2 + a_f^2)^{3/2}}(\zeta_b - \zeta_f)e^{-\frac{(\zeta_b - \zeta_f)^2}{a_b^2 + a_f^2}} \; , \nonumber \\
\zeta_{f,tt} &=& -4\alpha_f \zeta_f + \frac{4g_{fb}N_b}{\sqrt{\pi}(a_b^2 + a_f^2)^{3/2}}(\zeta_f - \zeta_b)e^{-\frac{(\zeta_b - \zeta_f)^2}{a_b^2 + a_f^2}} \label{ODE_for_COM} \; . \nonumber \\
\end{eqnarray}
We note that for small  $|\zeta_b - \zeta_f| \ll 1$ the above equations have the same form as the system (4)-(5) considered in~\cite{Ferrier}:
\begin{eqnarray}
\zeta_{b,tt} &=& -\Omega_b^2 \zeta_b  - K_b \zeta_f \; , \nonumber \\
\zeta_{f,tt} &=& -\Omega_f^2 \zeta_f  - K_f \zeta_b \; ,
 \label{ODEs}
\end{eqnarray}
where
\begin{equation}
\Omega_{b,f}^2 =4\alpha_{b,f} - K_{b,f},\  K_{b,f}=\frac{4g_{fb}N_{f,b}}{\sqrt{\pi}(a_b^2 + a_f^2)^{3/2}} \; . \label{COM_omega_b_f}
\end{equation}

\subsubsection{Total center of mass}

The total center of mass (COM) is given by (assuming the same atomic mass for the two components)
\begin{equation}
X(t)= \frac{ \int_{-\infty}^{\infty}  x \left( |\psi_b|^2 + |\psi_f|^2 \right) dx }{N_b + N_f} \; , \label{def_COM}
\end{equation}
which, within the Gaussian ansatz (\ref{VA}), takes the form
\begin{equation}
X(t) = \frac{N_b \zeta_b(t) + N_f \zeta_f(t)}{N_b + N_f} \; . \label{total_COM_ref}
\end{equation}
Forming the second-order derivative from Eqs.~(\ref{ODE_for_COM}) while keeping the widths $a_b$ and $a_f$ constant, we obtain $X_{tt} = -4(\alpha_b N_b \zeta_{b}(t) + \alpha_f N_f \zeta_{f}(t)) /(N_b + N_f)$.
An interesting limiting case is: $\alpha_b = \alpha_f=\alpha$, for which $X_{tt} = -4\alpha X$.
We can choose to center the trap $n=b,f$ at position $d_n$ that is to replace
$\alpha_n x^2 \psi_n$ by $\alpha_n (x-d_n)^2 \psi_n$ in Eq. (\ref{sys1}) for $t \geq 0$. In this case
the corresponding wave functions in Eq. (\ref{VA}) are centered around $d_n$, i.e. we have
the time averages $\langle \zeta_{n} \rangle = d_n$. So for
$\alpha_b = \alpha_f=\alpha$ Eq. (\ref{total_COM_ref}) have the form

\begin{equation}
X(t)  =  \frac{N_b \langle \zeta_{b} \rangle + N_f \langle \zeta_{f}\rangle}{N_b + N_f} (1-\cos(2\sqrt{\alpha} t))
\; . \label{COM_tot_X}
\end{equation}

\begin{figure}
\hspace{-5mm}
\includegraphics[scale=0.36]{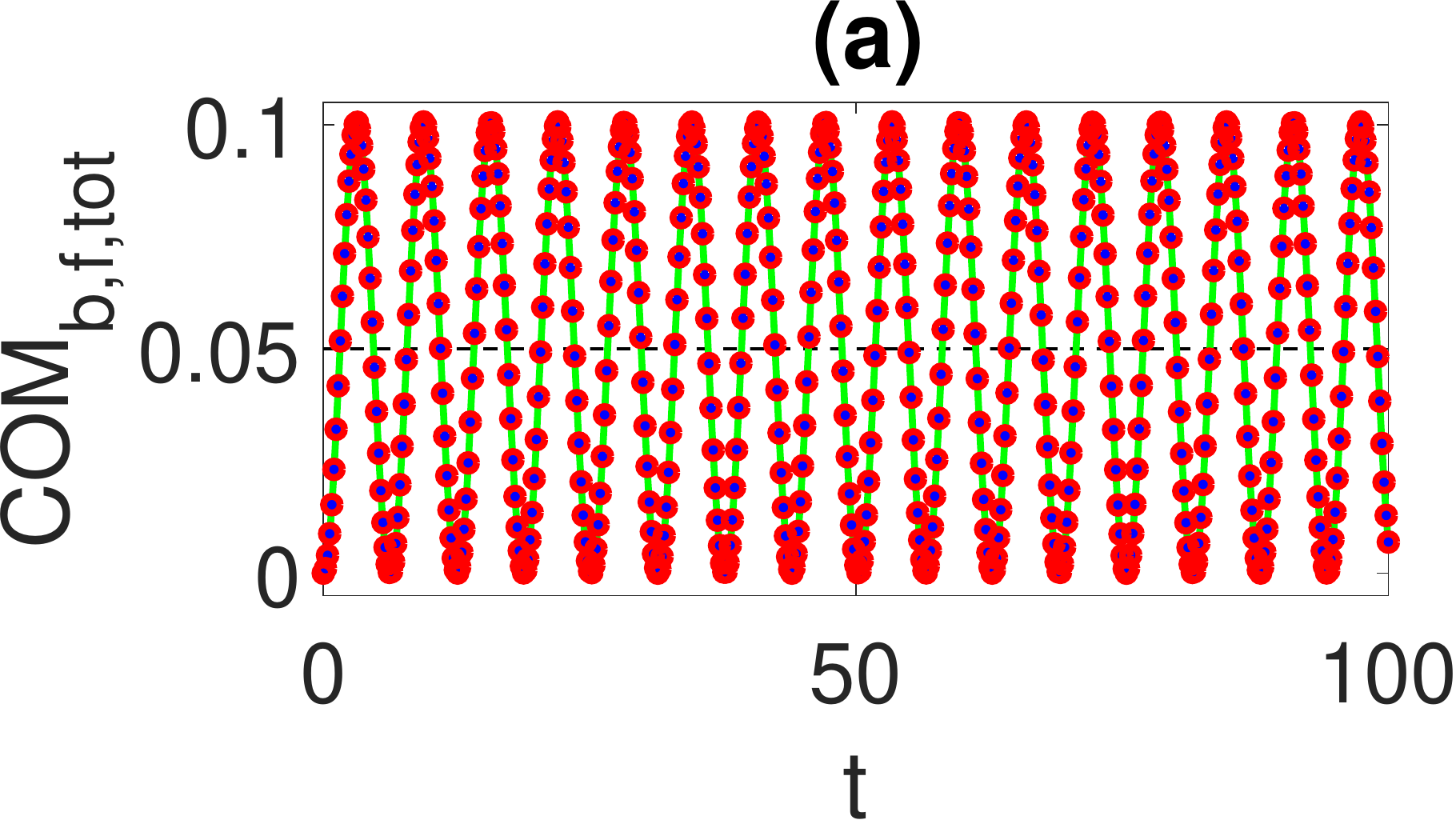} 
\includegraphics[scale=0.32]{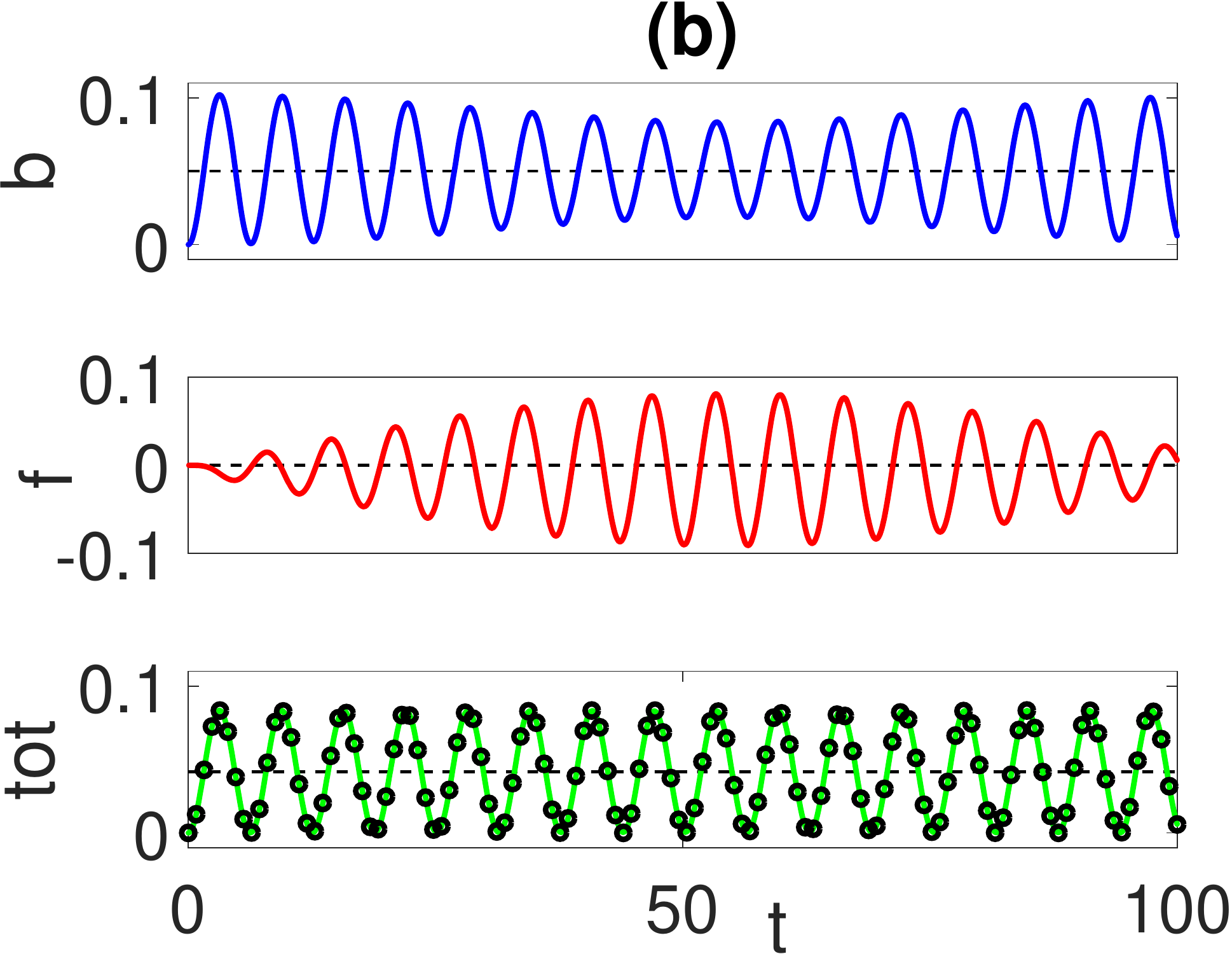} 
\caption{(Color online) Examples of dynamics of center of mass modes (COM) of the densities from the full PDE model~(\ref{sys1}).
Colors in use are: blue (b), red (f), and green (tot).
(a) At $t=0$ we have moved the trap for both uncoupled ($g_{fb}=0$) components by $d_b = d_f = 0.05$.
We find only one frequency (equals to unity in the present parameters, $\omega_0=2\sqrt{\alpha_b}=2\sqrt{\alpha_f}=1$) in all of the three signals $b$, $f$, and $tot$, plotted on top of each other.
(b) Coupled system ($g_{fb}=1/2$) where only the bosonic trap have been translated ($d_b=0.05$ and $d_f=0$) at $t=0$.
Using $N_b=10^3$ and $N_f=2 \cdot 10^2$, Eq.~(\ref{total_COM_ref}) gives the mean position $\langle X \rangle  \simeq 0.042$ for the total COM oscillations, shown by the dashed horizontal black line in the lower panel, while the black circles shows the results for the total COM oscillations from the analytic ODE based model Eq.~(\ref{COM_tot_X}).
As indicated by the beating patterns in (b) we find an additional frequency $\tilde{\omega}_n$ in both the components $b$ and $f$, but that cancels for the total density $tot$ (lower panel) from the numerical PDE solution, in agreement with Eq.~(\ref{COM_tot_X}).
See Fig.~\ref{fig_2} for quantitative results.
Physical parameters and the method for determining frequencies are discussed in Sec.~\ref{Sec_Numerical_simulations}.
\label{fig_1}}
\end{figure}

\noindent Hence, we note that unlike the COM of the respective components, the total COM
is not effected by the coupling parameter $g_{fb}$, see the lower panel of Fig.~\ref{fig_1}~(b).
Note that we use the initial conditions $\zeta_n(0)=0$ in Fig.~\ref{fig_1}, but as the centers of the
traps are displaced from zero this do not imply that the averages $\langle \zeta_{n} \rangle$, $n=b,f$, vanish.

\begin{figure}
\includegraphics[scale=0.35]{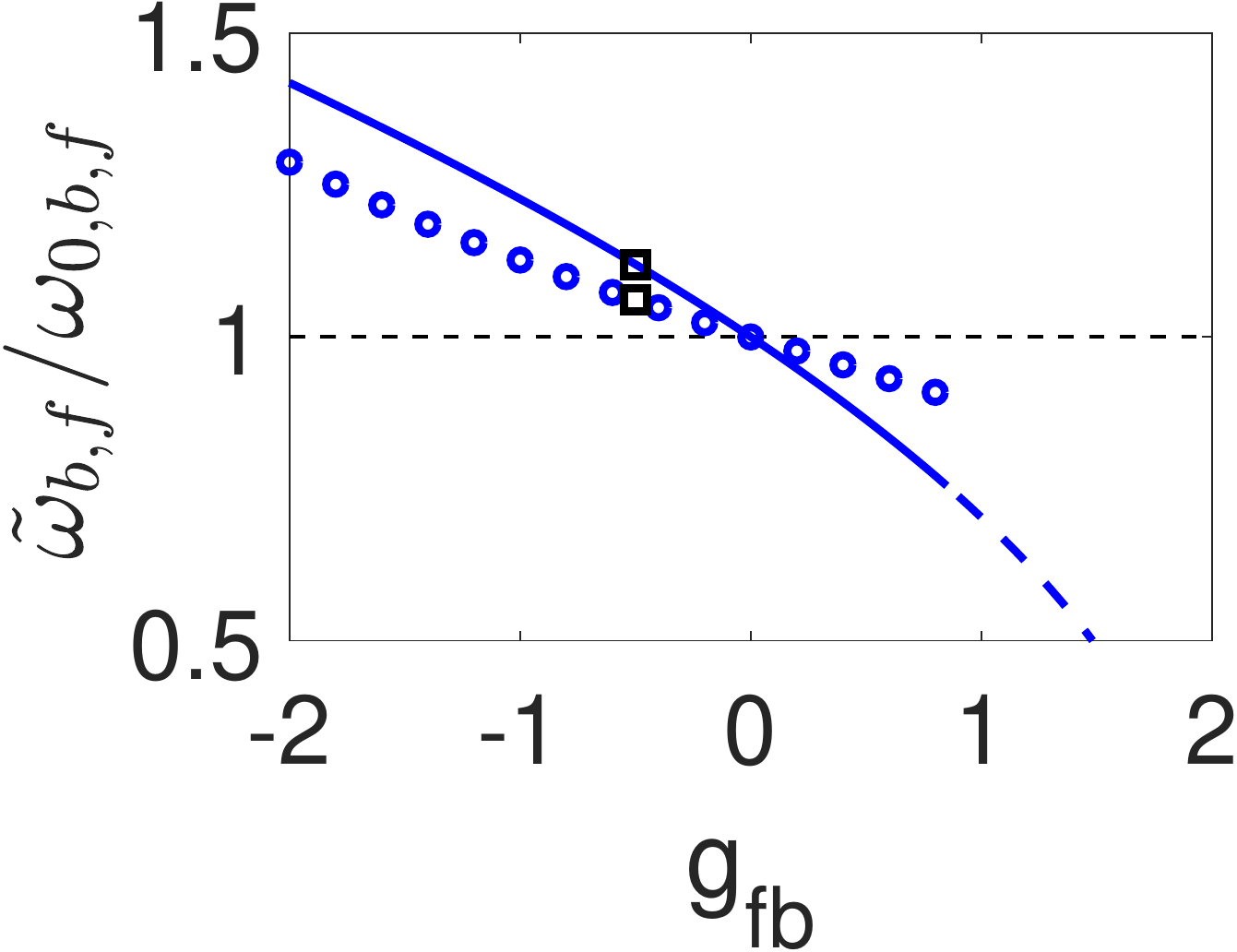}
\caption{(Color online) Center of mass frequency for different coupling $g_{fb}$. The frequency $\tilde{\omega}_{n}$ normalized with $\omega_{0,n}=2 \sqrt{\alpha_n}$ (equal unity with the parameters in use) is the same for the Bose- (b) and Fermi- (f) components, but not present in the total COM, see Eq.~(\ref{COM_tot_X}). Rings are from simulations of the full PDE model (\ref{sys1}), the solid (/dashed) curve shows the ODE based approximation~(\ref{ODE_tilde_COM}) with the widths $a_f=a_{f0}$ and $a_b=a_{b0}$ taken from Eqs.~(\ref{a_f_0_approx}) and~(\ref{a_b_0_approx}), respectively.
The black squares at $g_{fb}=-1/2$ corresponds to the black squares in Fig.~\ref{fig_4}.
Physical parameters and the method for determining frequencies are discussed in Sec.~\ref{Sec_Numerical_simulations}.
}
\label{fig_2}
\end{figure}

\subsubsection{Relative center of mass coordinate} \label{sec_Relative_center_of_mass_coordinate}

We define $\Delta\zeta= \zeta_b - \zeta_f$ and then, according to~(\ref{ODE_for_COM}) with $a_b$ and $a_f$ constant, we have for the case $|\Delta\zeta| \ll 1$, with $\alpha_b=\alpha_f=\alpha$
and  $N=N_b+N_f$:
\begin{equation}
\Delta\zeta_{tt}+4\alpha_R \Delta\zeta =0 \; , \label{COM_single_ODE}
\end{equation}
where
\begin{equation}
\alpha_R =\alpha - \frac{g_{fb} N}{\sqrt{\pi}(a_b^2+a_f^2)^{3/2}} \; . \label{alpha_R}
\end{equation}
So for $g_{fb}=0$ we have the frequency $\omega_{0}=\omega_{trap}=2\sqrt{\alpha}$ (equals unity in the parameters in use here).
For $g_{fb} \neq 0$ we have the frequency
\begin{equation}
\tilde{\omega} = 2 \sqrt{ \alpha - \frac{g_{fb} N}{\sqrt{\pi}(a_b^2+a_f^2)^{3/2}}}
\label{ODE_tilde_COM} \; .
\end{equation}
Due to the definitions~(\ref{total_COM_ref}) and $\Delta\zeta = \zeta_b - \zeta_f$, we see from Eq.~(\ref{ODE_tilde_COM}), that the Bose-coordinate $\zeta_b$ and the Fermi-coordinate $\zeta_f$ both have the two common frequences $\omega_{0,b,f}$ and $\tilde{\omega}_{b,f}$, see Figs.~\ref{fig_1} and~\ref{fig_2}.

\subsubsection{Parametric resonance for the center of mass mode}

\begin{figure}
\includegraphics[scale=0.4]{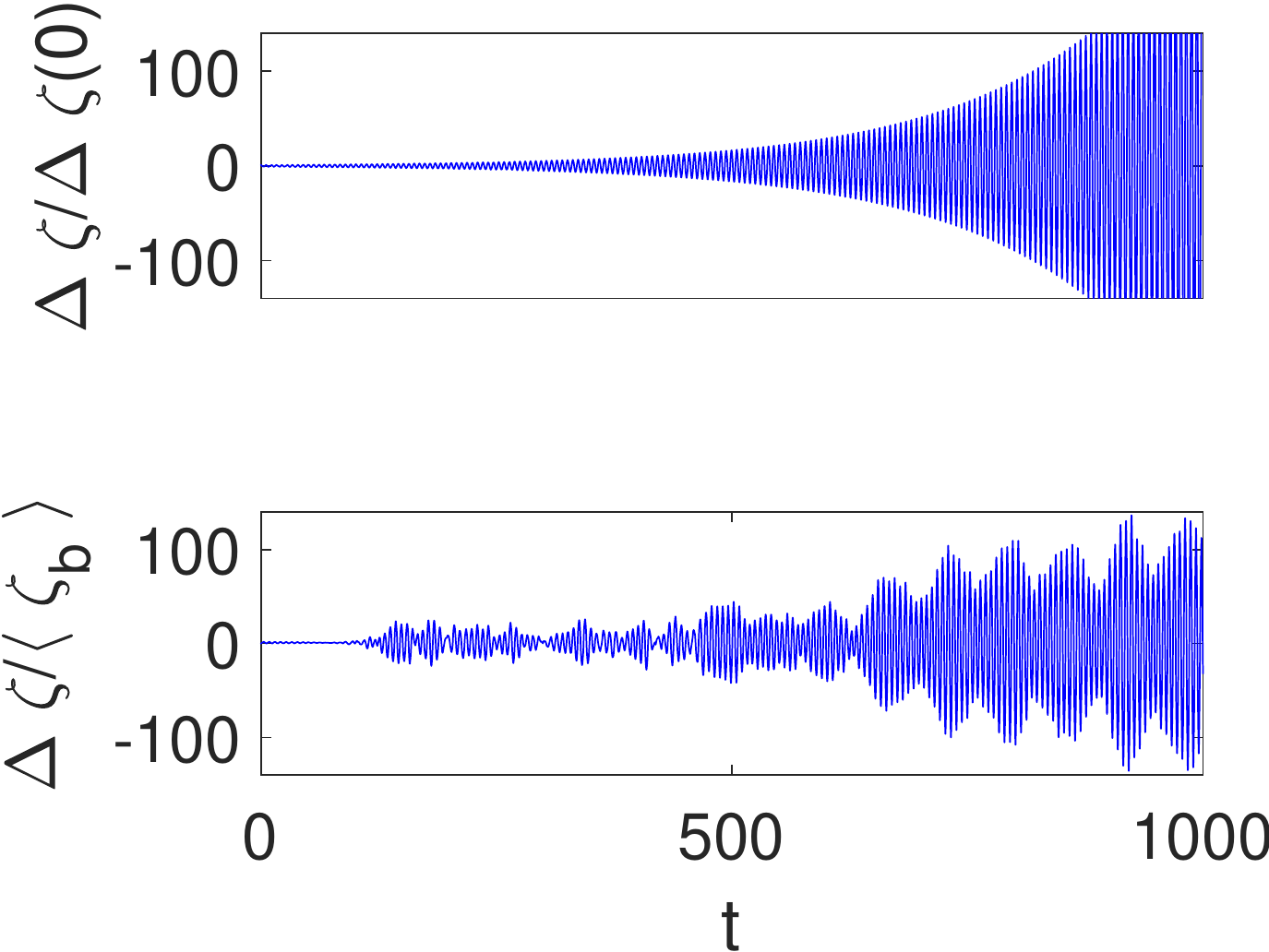}
\caption{Parametric resonance for the relative center of mass modes $\Delta \zeta(t)$.
Upper panel: Solution to the ODE model, from Eqs.~(\ref{Mathieu_equation}), (\ref{def_h}) and~(\ref{Omega_fb}), i.e. $\Omega_{fb}=2.2388$, with parameters $g_{fb}^{(0)}=-1/2$, $c_{fb}=0.1$, and the initial value $\Delta \zeta(0)=0.005$.
The widths $a_f=a_{f0}$ and $a_b=a_{b0}$ are taken from Eqs.~(\ref{a_f_0_approx}) and~(\ref{a_b_0_approx}) respectively.
Lower panel: Solution to the PDE model~(\ref{sys1}) with sinusoidal interspecies coupling~(\ref{TDgbANDg12}), with parameters $g_{fb}^{(0)}=-1/2$, $c_{fb}=0.05$, $\Omega_{fb}=2.1000$, and an initial translation of the trap for the Bose-component such that $d_b=0.005$ (see text).
Physical parameters are discussed in Sec.~\ref{Sec_Numerical_simulations}.
\label{fig_3}}
\end{figure}

Note that the oscillations of the relative distance between centers of the fermionic and bosonic clouds depends on the fermion-boson scattering length, while the motion of the total center of mass of the system do not depend on $g_{fb}$ (see above).
Experiments can use this dependence to study the coupling between the dynamics of bosonic and fermionic superfluids.
In particular if we consider the periodic modulations in time for the fermion-boson scattering length $g_{fb}$ according to Eq.~(\ref{TDgbANDg12}), we obtain for $\Delta\zeta$ the differential equation
\begin{equation}
\Delta\zeta_{tt} +\omega_{\zeta}^2 \left[ 1-h\sin \left(\Omega_{fb}t \right) \right]\Delta\zeta =0 \; , \label{Mathieu_equation}
\end{equation}
where
\begin{equation}
\omega_{\zeta}=2\sqrt{\alpha_R} \; ,  \;   \;  h=\frac{g_{fb}^{(0)}N c_{fb}}{\alpha_R \sqrt{\pi}(a_b^2 + a_f^2)^{3/2}} \; , \label{def_h}
\end{equation}
with $\alpha_R$ taken from~(\ref{alpha_R}).
This is the well known Mathieu equation.
Such that when
\begin{equation}
\Omega_{fb}=2\omega_{\zeta} \; ,   \label{Omega_fb}
\end{equation}
we have a first parametric resonance in the oscillations of the relative distance between the clouds (upper panel of Fig.~\ref{fig_3}).
The region of instability is the interval $-|h|\omega_{\zeta}/2< \epsilon< |h|\omega_{\zeta}/2$, with $h$ from~(\ref{def_h}), around the frequency $2\omega_{\zeta}$, that is $\Omega_{fb}=2 \omega_{\zeta}+\epsilon$.
For the full PDE model~(\ref{sys1}) parametric resonances can be found for driving frequences close to the prediction~(\ref{Omega_fb}), see the lower panel of Fig.~\ref{fig_3}.

\subsubsection{Different trap frequencies}
The frequencies of the normal modes to~(\ref{ODEs}) are
\begin{equation}
\Omega_{1,2}^2=\frac{\Omega_b^2 +\Omega_f^2}{2} \mp \sqrt{\frac{(\Omega_b^2 - \Omega_f^2)^2}{4} + K_b K_f} \; .
\end{equation}
An explicit expression is given by
$$
\Omega_{1,2}^2=2(\alpha_b + \alpha_f) - \frac{K_b + K_f}{2}
$$
\begin{equation}
 \mp \sqrt{ 4\left( \alpha_b - \alpha_f \right)^2 + 2\left( \alpha_b - \alpha_f \right) \left( K_f - K_b \right) + \frac{ \left( K_b + K_f \right)^2 }{4}} \; . \label{explicit_COM_frequencies}
\end{equation}

\begin{figure}
\includegraphics[scale=0.4]{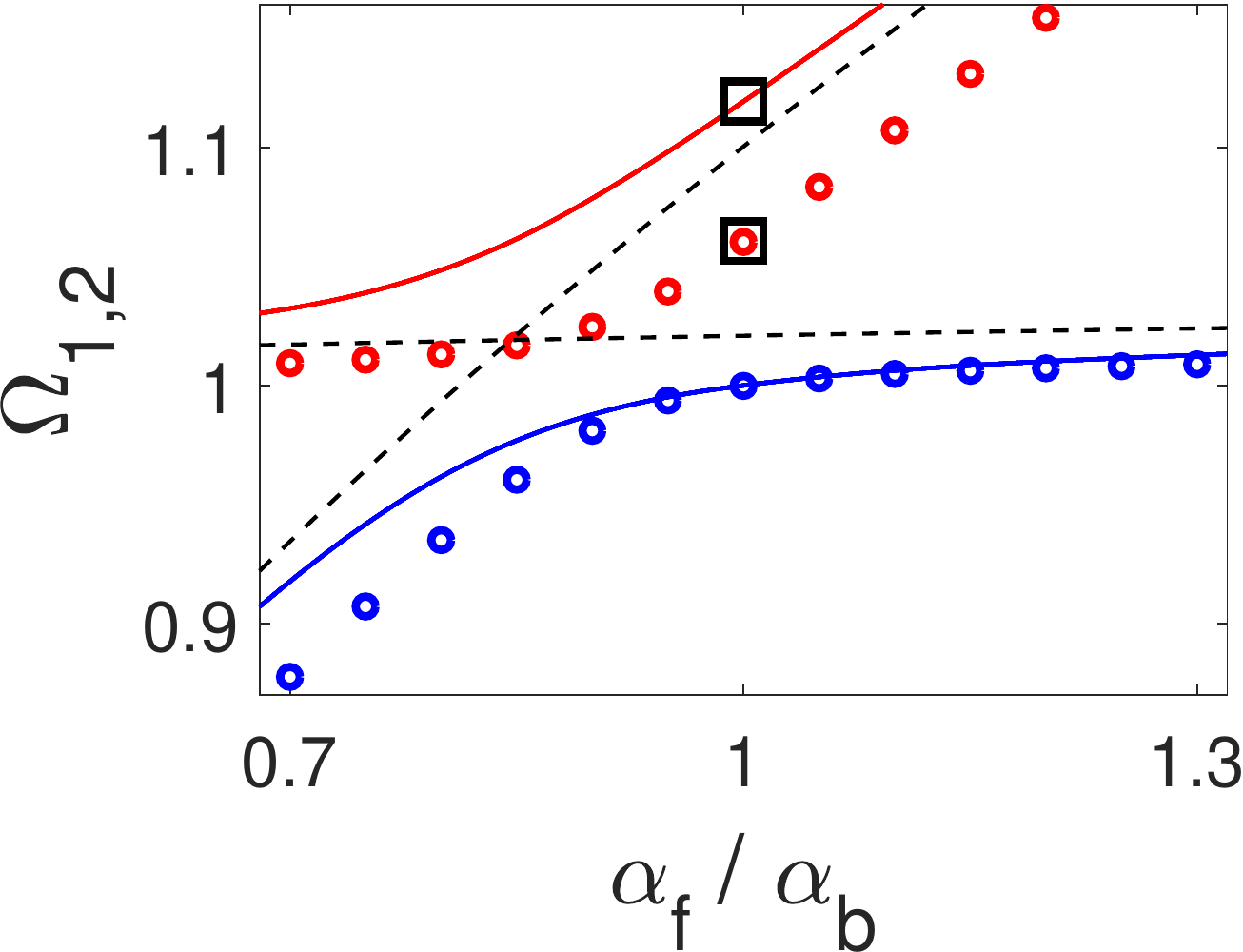} 
\caption{(Color online) Center of mass frequencies $\Omega_{1,2}$ for different traps $\alpha_b \neq \alpha_f$.
Blue ($\Omega_1$, lower rings) and red ($\Omega_2$, upper rings) are from simulations of the full PDE model~(\ref{sys1}), the solid curves shows the ODE based approximation~(\ref{explicit_COM_frequencies}) with the widths $a_f=a_{f0}$ and $a_b=a_{b0}$ taken from Eqs.~(\ref{a_f_0_approx}) and~(\ref{a_b_0_approx}) respectively.
The dashed lines show the $K_{b,f} \ll |\alpha_b-\alpha_f|$ limits of Eq.~(\ref{COM_small_K_limit}).
Since $g_{fb}=-1/2$ here, the black squares at $\alpha_b=\alpha_f$ corresponds to the data points in the black squares of Fig.~\ref{fig_2}.
The top black square is also the $|\alpha_b-\alpha_f| \ll K_{b,f}$ limiting result for $\Omega_2$ of Eq.~(\ref{COM_large_K_limit}), while $\Omega_1=1$.
Physical parameters and the method for determining frequencies are discussed in Sec.~\ref{Sec_Numerical_simulations}.
}
\label{fig_4}
\end{figure}

\noindent In Fig.~\ref{fig_4} we compare the prediction for the frequencies $\Omega_{1,2}$ from Eq.~(\ref{explicit_COM_frequencies}) with simulations of the full PDE model~(\ref{sys1}). The figure shows
that the relative deviation between full numerical solutions and the prediction from
Eq.~(\ref{explicit_COM_frequencies}) is less than about $5.4\%$. This is a good agreement supporting
the validity of our collective coordinate approach in section \ref{sec_The_model}.

From Eq.~(\ref{explicit_COM_frequencies}) we also have the two important limiting cases from before for the following parameters:

\paragraph{Small $|\alpha_b -\alpha_f| \ll K_{b,f}$}
Then we have from~(\ref{explicit_COM_frequencies}) the frequencies
\begin{equation}
\Omega_{1,2}^2 \simeq  2 \left( \alpha_b + \alpha_f \right) - \frac{K_b+K_f}{2} \mp \frac{K_b+K_f}{2} \; .
\end{equation}
Hence, we have the two limiting frequences ($\alpha = \alpha_b = \alpha_f$): $\Omega_{1}= 2 \sqrt{\alpha}$; and ($N = N_b + N_f$)
\begin{equation}
\Omega_{2}= 2 \sqrt{\alpha - \frac{K_b+K_f}{4} }  = 2 \sqrt{\alpha - \frac{g_{fb}N}{\sqrt{\pi} ( a_b^2 + a_f^2 )^{3/2} } } =  \tilde{\omega} \; . \label{COM_large_K_limit}
\end{equation}
We note that the above limit agree with the results of Sec.~\ref{sec_Relative_center_of_mass_coordinate}, see e.g. Eq.~(\ref{ODE_tilde_COM}) and Fig.~\ref{fig_2};

\paragraph{Small $K_{b,f} \ll |\alpha_b - \alpha_f|$}
For which we get from~(\ref{explicit_COM_frequencies}) the frequencies
\begin{eqnarray}
\Omega_{1}^2 \simeq 4 \alpha_{b} -K_b \; ,\nonumber\\
\Omega_{2}^2 \simeq 4 \alpha_{f}-K_{f} \; ,\label{COM_small_K_limit}
\end{eqnarray}
i.e., the frequences $\Omega_{b,f}$ from Eq.~(\ref{COM_omega_b_f}), see Fig.~\ref{fig_4}.

\subsubsection{Parametric resonance for the COM with different trap frequencies} % \label{sec_...}
In the case of a time-dependent fermion-boson scattering length, see Eq.~(\ref{TDgbANDg12}), the system~(\ref{ODEs}) represents two coupled Mathieu equations
\begin{eqnarray}
\zeta_{b,tt} + 4\alpha_b \zeta_b &=& + K_b(t) \Delta \zeta \; , \nonumber \\
\zeta_{f,tt} + 4\alpha_f \zeta_f &=& - K_f(t) \Delta \zeta \; , %%% (t) ???
 \label{COM_coupled_ODEs}
\end{eqnarray}
which in the case of $\alpha_b=\alpha_f$ leads to Eq.~(\ref{COM_single_ODE}) in Sec.~\ref{sec_Relative_center_of_mass_coordinate}.

\begin{figure}
\includegraphics[scale=0.45]{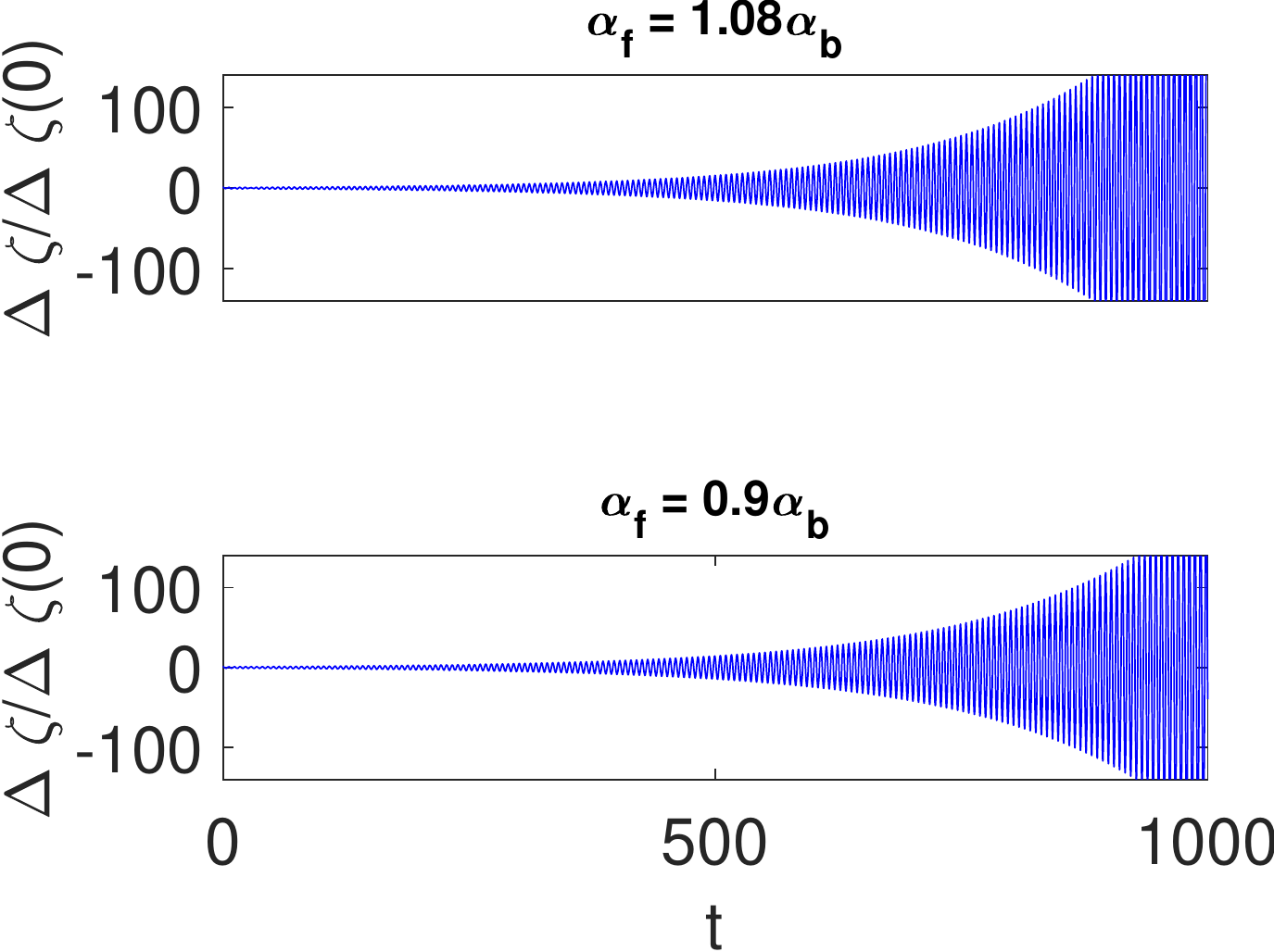}
\caption{%PREL!?
Parametric resonance for the relative center of mass modes $\Delta \zeta(t)$ for different traps $\alpha_b \neq \alpha_f$ within the ODE model.
Upper panel: Solution to the ODE model, Eq.~(\ref{COM_coupled_ODEs}), with the driving $\Omega_{fb}=2 \Omega_2= 2.2638$, with parameters $g_{fb}^{(0)}=-1/2$, $c_{fb}=0.1$, $\alpha_b=1/4$, $\alpha_f=1.08 \alpha_b$, and the initial value $\Delta \zeta(0)=0.005$.
The widths $a_f=a_{f0}$ and $a_b=a_{b0}$ are taken from Eqs.~(\ref{a_f_0_approx}) and~(\ref{a_b_0_approx}) respectively.
Lower panel: Solution to the ODE model, Eq.~(\ref{COM_coupled_ODEs}), with the driving $\Omega_{fb}=2 \Omega_2= 2.2152$, with the same parameters as in the upper panel except that $\alpha_f=0.9 \alpha_b$.
No qualitative differences are seen for the $\alpha_b \neq \alpha_f$ case compared to Fig.~\ref{fig_3}, that shows both ODE and PDE results for the case $\alpha_b = \alpha_f$.
Physical parameters are discussed in Sec.~\ref{Sec_Numerical_simulations}.
\label{fig_5}}
\end{figure}

As is showed in the work~\cite{Hansen}, parametric resonances in oscillations are possible at the following driving conditions in Eq.~(\ref{COM_coupled_ODEs}):
\begin{equation}
\Omega_{fb} = 2\Omega_1;  \ 2\Omega_2; \ |\pm \Omega_1 \pm \Omega_2| \; .
\end{equation}
See Fig.~\ref{fig_5} for two numerical examples.

Let us finally estimate possible experimental parameters.
The parameters for the system $^{6}$Li-$^{7}$Li at the conditions of the experiment~\cite{Ferrier} are:
the fermion-boson scattering length is $a_{fb}=40a_0$;
the trap frequencies are $\omega_{trap,b}=2\pi \cdot 15.2$ Hz and $\omega_{trap,f} = 2\pi \cdot 16.8$ Hz;
with the transverse frequency $\omega_{\perp}= 550$ Hz, such that the transverse length is $l_{\perp} \simeq 4$ $\mu$m ;
the number of  fermionic $^{6}$Li-atoms is $N_f=3.5\cdot 10^5$ and the number of bosonic $^{7}$Li-atoms is $N_b = 4 \cdot 10^4$;
the widths of the bosonic and fermionic clouds are $a_b \simeq 2.5l_{\perp}$ and $a_f \simeq 10 l_{\perp}$;
the coupling parameter in~(\ref{ODEs}), for the trapping parameters $\alpha_b=0.001$  and $\alpha_f = 0.0011$ , are $K_b \simeq 0.002$ and $K_f \simeq 10 K_b$.

Then we can conclude that the frequency of the bosonic center of mass is reduced due to the fermi-bose interaction from $15.3$ Hz, to $14.4$ Hz, while the fermionic is not practically changed. This agree with the experimental observation reported in~\cite{Ferrier}.

The atomic scattering length dependence on the external magnetic field $B$ in the region of the Feshbach resonance is:
\begin{equation}
a_{fb} = a_{bg} \left( 1-\frac{\Delta}{B_0 - B(t)} \right) \; ,
\end{equation}
where $B_0$ is the Feshbach resonance position, $\Delta $ is the width of the resonance, and $a_{bg}$ is the background atomic scattering length.
For the $^{6}$Li-$^{7}$Li mixture with $B_0=360$~G~\cite{Abeelen,Kempen} the atomic scattering length $a_{fb}$ can be varied by applying the external magnetic field varying periodically in the time near the Feshbach resonance at $B_0$, $B(t)= B_{av} + B_1\sin(\Omega t)$, where $B_{av}$ is the averaged value of the field.

Then for the frequencies of the parametric resonances in dimension variables we find:
$2\Omega_1 = 2\pi \cdot 31$ Hz and $2\Omega_2 = 2\pi \cdot 32.1$ Hz.

\subsection{Breathing modes}

For the width of the bosonic subsystem we have from~(\ref{Euler-Lagrange-Equation}) the equation (see the Appendix for details)
\begin{eqnarray}
a_{b,tt} = \frac{4}{a_b^3}  + \frac{\sqrt{2}g_b N_b}{\sqrt{\pi}a_b^2} - 4\alpha_b a_b
-4 N_f g_{fb} \frac{dF}{d a_b} \; , \label{Eq_a_b_tt}
\end{eqnarray}
where
$$
 F = \frac{ e^{-\frac{(\zeta_b - \zeta_f)^2}{a_b^2 + a_f^2}} }{\sqrt{\pi(a_b^2 + a_f^2 )}} \; .
$$

\noindent For the width of the fermionic subsystem we obtain the equation
\begin{equation}
a_{f,tt} = \frac{4}{a_f^3} - 4\alpha_f a_f
+\frac{8\pi \kappa N_f^2}{3\sqrt{3}a_f^3}
- 4 N_b g_{fb} \frac{dF}{da_f} \; . \label{Eq_a_f_tt}
\end{equation}

\noindent The equilibrium values for the widths of bosonic and fermionic clouds, from here on denoted $a_{b0}$ and $a_{f0}$, are given by the solutions to the equations
\begin{eqnarray}
\frac{4}{a_{b}^3}  + \frac{\sqrt{2}g_b N_b}{\sqrt{\pi}a_{b}^2} - 4\alpha_b a_{b}
- 4 N_f g_{fb} \frac{dF}{d a_b} &=& 0 \; , \nonumber \\
  \frac{4}{a_{f}^3} - 4\alpha_f a_{f}
+\frac{8\pi \kappa N_f^2}{3\sqrt{3}a_{f}^3} - 4 N_b g_{fb} \frac{dF}{da_f} &=& 0 \; . \label{equilibrium_widths}
\end{eqnarray}
We can find the frequencies of breathing modes by a linearization of Eqs. (\ref{Eq_a_b_tt}) and (\ref{Eq_a_f_tt}) near  the fixed point $(a_{b0}, \: a_{f0})$ by the substitutions $a_b = a_{b0} + \delta a_b$ and $a_f =a_{f0}+\delta a_f $.
From the equations for $\delta_b$ and $\delta_f$ we then obtain the following coupled system of equations
\begin{eqnarray}\label{corr1}
\delta a_{b,tt} + \omega_b^2 \delta a_b = -\epsilon_1 (t) \delta a_f \; , \nonumber \\
\delta a_{f,tt} + \omega_f^2 \delta a_f = -\epsilon_2 (t) \delta a_b \; , \label{Eq_lin_a_b_a_f}
\end{eqnarray}
where
\begin{eqnarray}
\omega_b^2 = \frac{12}{a_{b0}^4} \left( 1 + \frac{g_b N_b a_{b0}}{ 3\sqrt{2\pi}} \right) + 4\alpha_b + 4 N_f g_{fb}\frac{d^2 F}{d a_b^2}|_{a_b=a_{b0}} \; , \nonumber \\
\omega_f^2 = \frac{12}{a_{f0}^4} \left( 1 + \frac{2\pi \kappa N_f^2}{3\sqrt{3}} \right) + 4\alpha_f + 4N_b g_{fb} \frac{d^2 F}{d a_{f}^2}|_{a_f=a_{f0}} \; , \nonumber \\ \label{omega_b_and_omega_f}
\end{eqnarray}
and
\begin{eqnarray}
\epsilon_1 &=& 4N_f g_{fb} (t) \frac{d^2 F}{da_b da_f}|_{a_{b,f}=a_{b0,f0}} \; , \nonumber \\
\epsilon_2 &=&  4N_b g_{fb} (t) \frac{d^2 F}{da_b da_f}|_{a_{b,f}=a_{b0,f0}} \; . \label{epsilon_1_and_epsilon_2}
\end{eqnarray}
The time dependence in $\epsilon_{1,2}$ comes into play when the interspecies coupling $g_{fb}$ is time dependent (e.g.) according to~Eq.~(\ref{TDgbANDg12}).
For Eq. (\ref{Eq_lin_a_b_a_f}) we then have that the frequencies of the normal modes for the coupled bosonic and fermionic systems are
\begin{equation}
\omega_{1,2}^2 = \frac{\omega_b^2 + \omega_f^2}{2} \mp \sqrt{\frac{(\omega_f^2 -\omega_b^2)^2}{4}+ \epsilon_1\epsilon_2} \; . \label{omega_1_and_omega_2}
\end{equation}

To compare the breathing mode frequences given by Eq.~(\ref{omega_1_and_omega_2}) against full PDE simulations we study some special cases.

\subsubsection{Breathing modes for uncoupled components ($g_{fb}=0$)}

First we solve the system~(\ref{equilibrium_widths}), which is now uncoupled, and obtain the equilibrium value for the widths of the fermionic subsystem $a_{f0} = \left(   ( 3\sqrt{3} + 2\pi \kappa N_f^2)/(3\sqrt{3}\alpha_f) \right)^{1/4}$.
In particular we see that the $N_f \rightarrow \infty$ limit
\begin{equation}
a_{f0} \rightarrow \left( \frac{ 2\pi \kappa}{3\sqrt{3} \alpha_f} \right)^{1/4}  \sqrt{N_f} \; , \label{a_f_0_approx}
\end{equation}
compares with the Thomas-Fermi width~\cite{Thomas-Fermi}
except for a numerical factor $ \left( \pi /( 6 \sqrt{3} ) \right)^{1/4}  \simeq 0.74$.

Now let us obtain an analytic expression also for $a_{b0}$ by neglecting the term $4/a_b^3$ in the upper line of~(\ref{equilibrium_widths}), which can be motivated for large values of $N_b$ and/or $a_b$.
Then we have the real positive solution
\begin{equation}
a_{b0}  = \left( \frac{ g_b }{2 \sqrt {2 \pi } \alpha_b } N_b \right)^{1/3} \; , \label{a_b_0_approx}
\end{equation}
which compares with the Thomas-Fermi width~\cite{Thomas-Fermi}
except for a numerical factor $ \left(2/(9\pi) \right)^{1/6} \simeq 0.64$.

Now we take the (simplified) analytic expressions~(\ref{a_f_0_approx}) and~(\ref{a_b_0_approx}) for the equilibrium widths and insert them into~(\ref{omega_b_and_omega_f}), to obtain
\begin{eqnarray}
\omega_b^2 = \frac{12}{a_{b0}^4} \left( 1 + \frac{g_b N_b a_{b0}}{ 3\sqrt{2\pi}} \right) + 4\alpha_b \simeq   \frac{4 g_b N_b }{ \sqrt{2\pi}} \frac{1}{a_{b0}^3} + 4\alpha_b \; , \nonumber \\
\omega_f^2 = \frac{12}{a_{f0}^4} \left( 1 + \frac{2\pi \kappa N_f^2}{3\sqrt{3}} \right) + 4\alpha_f \simeq  \frac{ 8 \pi \kappa N_f^2}{\sqrt{3}} \frac{1}{a_{f0}^4} + 4\alpha_f \; , \nonumber \\
\end{eqnarray}
which gives $\omega_b^2 =  12 \alpha_b$ and $\omega_f^2 =  16 \alpha_f$.
Now with $\alpha_n=\frac{1}{2} m_n \omega_{trap,n}^2$, and $m_n =1/2$, we finally have
\begin{equation}
\omega_b^2 =  3 \omega_{trap,b}^2 , \ \omega_f^2 =  4 \omega_{trap,f}^2 \; . \label{omega_b_and_omega_f_case_1}
\end{equation}
This agrees with the result in the literature for Thomas-Fermi based models \cite{MenottiPRA2002,FuchsPRA2003}, and hence is an alternative derivation for the breathing mode frequency of each (uncoupled) component.

\begin{figure}
\includegraphics[scale=0.23]{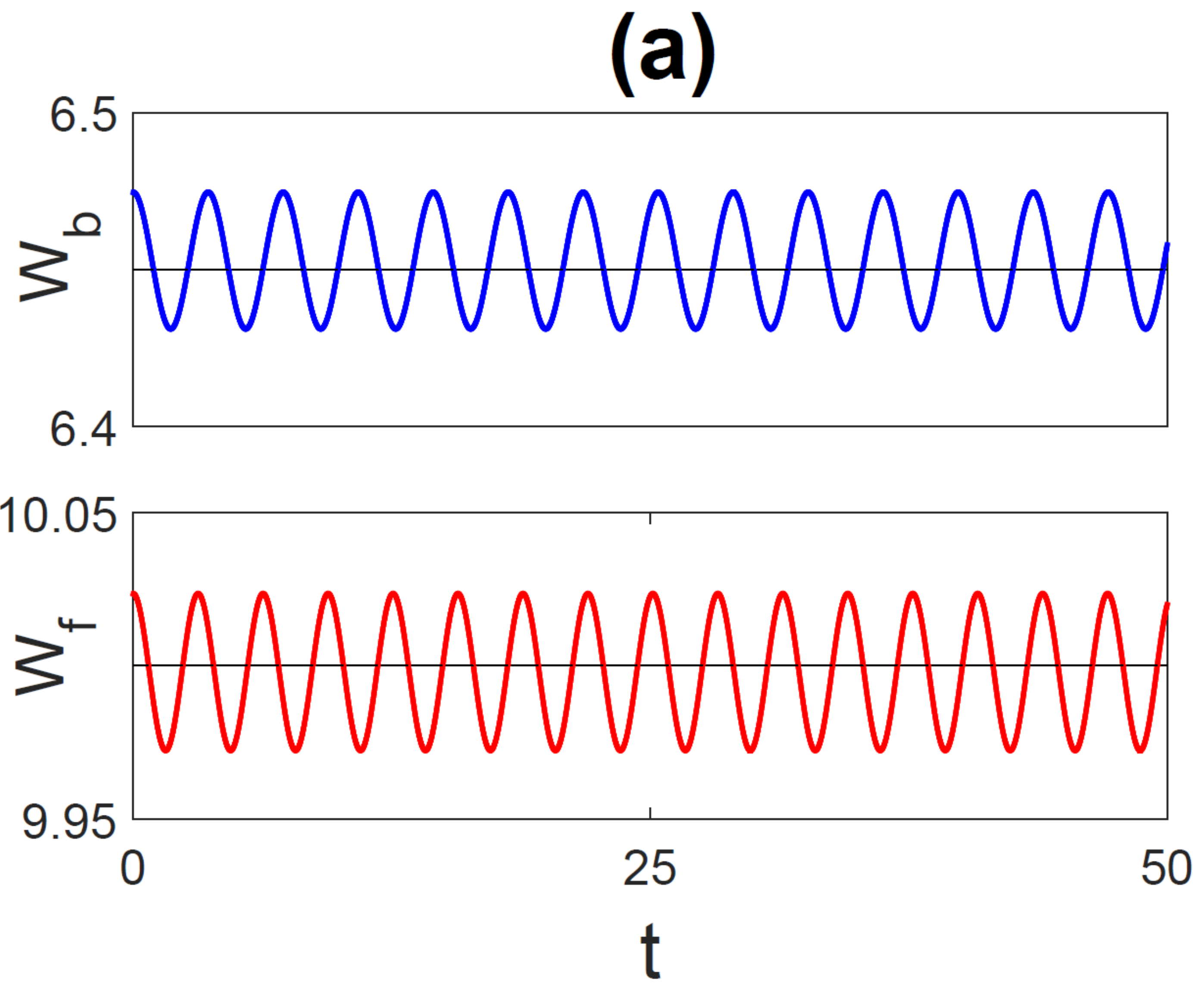} 
\includegraphics[scale=0.37]{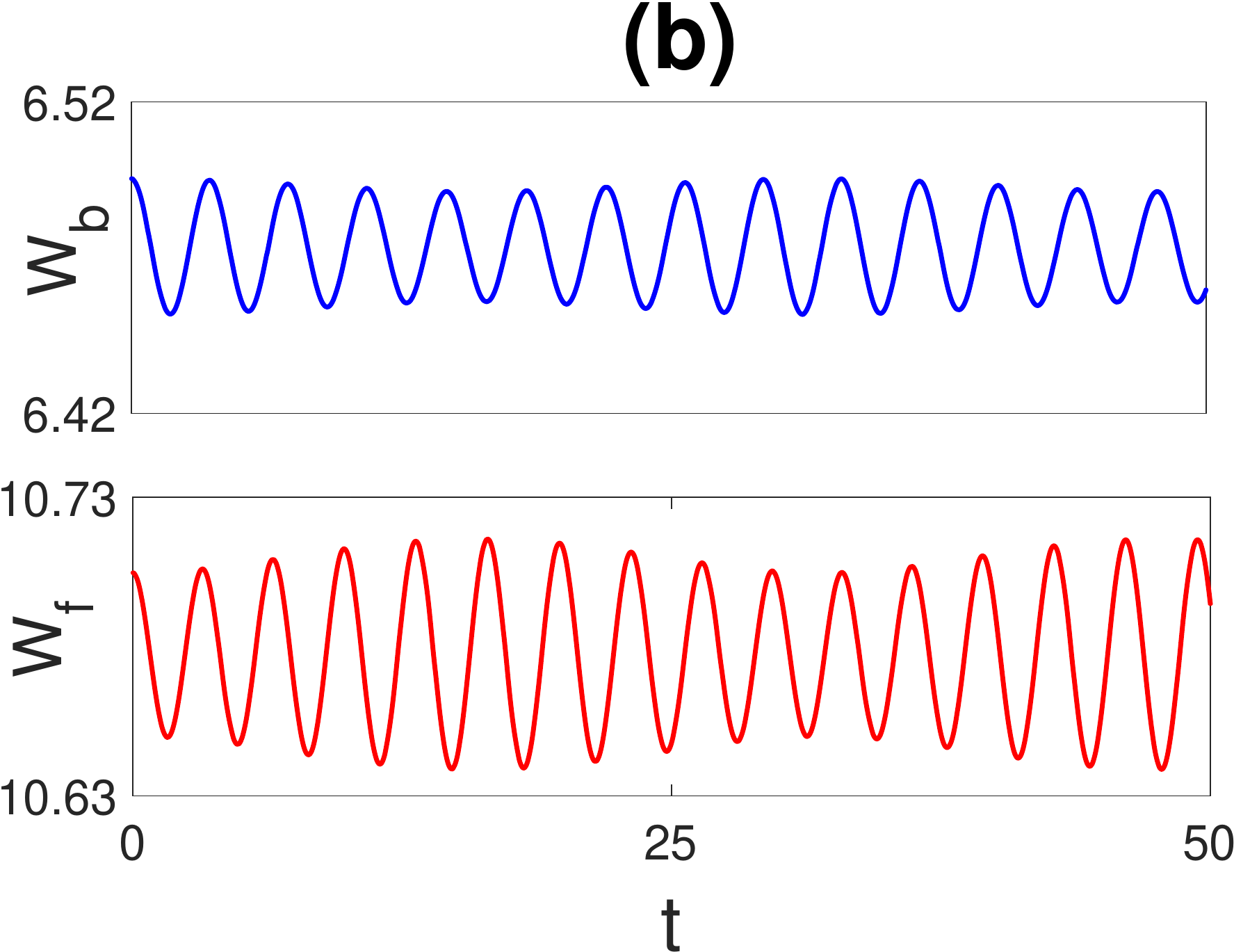}  
\caption{(Color online) Examples of breathing mode dynamics. We plot the RMS widths of the densities (anticorrelated to the peak density) of the two components, from the full PDE model~(\ref{sys1}). (a) is for the uncoupled system. (b) is for the coupled system ($g_{fb}=1/2$).
The straight lines in (a) shows the RMS width of the Thomas-Fermi profiles~\cite{Thomas-Fermi}.
We determined the single frequencies for the RMS signal in each component in (a) to $\omega_b=1.732$ and $\omega_f=2.000$, which is in agreement with Eq.~(\ref{omega_b_and_omega_f_case_1}).
As indicated by the beating patterns in (b), we then have two frequencies for each RMS signal, the same in both component, which are in agreement with Eq.~(\ref{omega_1_and_omega_2}). See Fig.~\ref{fig_7} for quantitative results.
Physical parameters and the method for determining frequencies are discussed in Sec.~\ref{Sec_Numerical_simulations}.
\label{fig_6}}
\end{figure}

In Fig.~\ref{fig_6}~(a), we illustrate this case, that was used to check the accuracy of the numerical procedures.

\subsubsection{Breathing modes for constant $g_{fb}\neq 0$ (not coupled to COM modes i.e. $\zeta_{b}=\zeta_{f}=0$)}

Also for the breathing modes the two same frequencies occurs in both components when there is a coupling $g_{fb} \neq 0$.
%But unlike for the COM modes where one of the frequencies always remains unity ($\omega_{0,n}=1$ for $\alpha_b=\alpha_f=1/4$)
Now we have two frequencies $\omega_{1,2}$ that both differs from the respective uncoupled values given by~(\ref{omega_b_and_omega_f_case_1}): $\omega_b=\sqrt{3}\omega_{trap,b}$ and $\omega_f=2\omega_{trap,f}$ ($\omega_{trap,n}=2\sqrt{\alpha_n}$ both equals unity in the PDE simulations here), see Fig.~\ref{fig_6}~(b).

\begin{figure}
\includegraphics[scale=0.3]{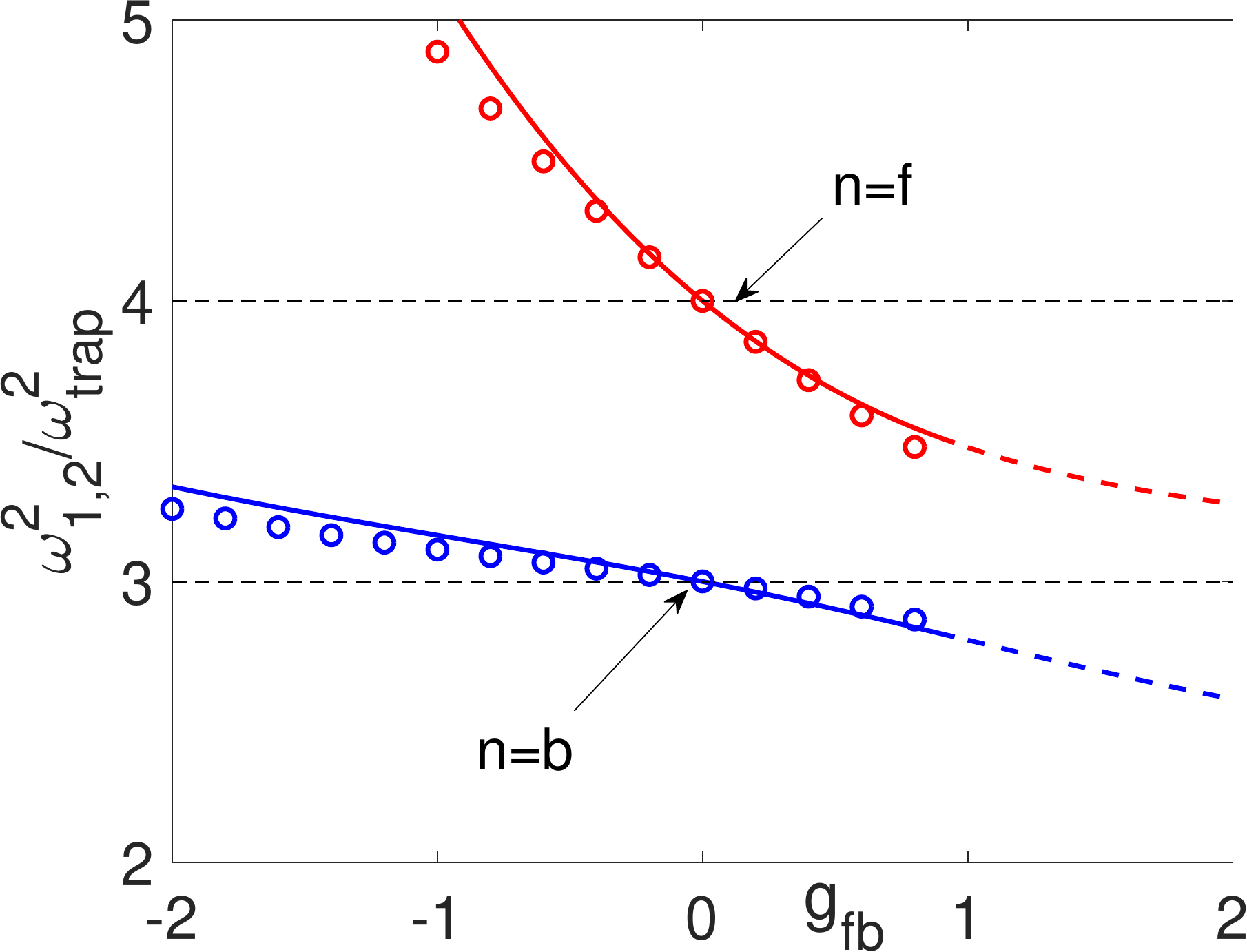}
\caption{(Color online) Breathing mode frequencies for different couplings $g_{fb}$.
The solid (/dashed) curves shows the predictions from the ODE model, from Eqs.~(\ref{omega_b_and_omega_f}), (\ref{epsilon_1_and_epsilon_2}) and (\ref{omega_1_and_omega_2}). Rings shows results from the full PDE simulations. For the PDE calculations the initial state was chosen as the groundstate for $\omega_{trap,n}=0.995$, i.e., by about 1$\%$ lower values of $\alpha_n$ compared to the evolution in real time.
Physical parameters and the method for determining frequencies are discussed in Sec.~\ref{Sec_Numerical_simulations}.
\label{fig_7}}
\end{figure}

We now solve the coupled system~(\ref{equilibrium_widths}) for real positive roots (numerically) for each value of $g_{fb}$ to obtain the equilibrium values for the widths of bosonic and fermionic subsystems.
Hence, by doing this we can evaluate Eqs.~(\ref{omega_b_and_omega_f}), (\ref{epsilon_1_and_epsilon_2}) and~(\ref{omega_1_and_omega_2}) numerically for different values of the coupling $g_{fb}$,
see Fig.~\ref{fig_7},
and/or for different trap strengths $\alpha_b$ and $\alpha_f$.

\subsubsection{Parametric resonance of breathing modes}

For the periodic modulations in time of the scattering length $a_{s,fb}$ i.e. with $g_{fb}$ according to Eq.~(\ref{TDgbANDg12}), the system~(\ref{corr1}) is two coupled Mathieu like equations.
This system has been investigated in the work~\cite{Hansen}.
Applying the results to our case (with $\zeta_{b}=\zeta_{f}=0$) we accordingly expect resonances in the breathing mode oscillations.
The resonances can occur for a driving frequency $\Omega_{fb}$ near twice of the normal modes eigenfrequencies~(\ref{omega_1_and_omega_2}) (and their subharmonics) $2\omega_n$, and also close to the combination frequencies
\begin{equation}
\Omega_{fb} = |\pm\omega_{1} \pm \omega_{2}| \; . \label{BM_driving_frequency}
\end{equation}
For small $\epsilon_{1,2}\ll \omega_{b,f} $ we have the following estimates from~(\ref{omega_1_and_omega_2})
$$
\omega_{1,2} \simeq \omega_{b,f} \left(1 \mp \frac{\epsilon_1\epsilon_2}{2\omega_{b,f}^2(\omega_f^2 - \omega_b^2)} \right) \; ,
$$
such that
$$
\omega_1 \pm \omega_2 \simeq (\omega_b \pm \omega_f) \left(1 \mp \frac{\epsilon_1\epsilon_2}{2\omega_b\omega_f(\omega_b + \omega_f)^2} \right) \; .
$$
As was shown with a multiscale analysis in the Appendix of~\cite{Abdullaev2013} $|\omega_1-\omega_2|$ is stable.
Furthermore, the driving frequency $\omega_1+\omega_2$ have a higher gain than $2\omega_n$ and is therefore used in the numerical examples of parametric resonance in the breathing modes presented in Figs.~\ref{fig_8} and~\ref{fig_9}. In the Appendix of~\cite{Abdullaev2013} also the region of instability for $\omega_1+\omega_2$ was determined.
The exponentially increasing widths of the Bose and Fermi clouds demonstrate the instability of this resonance.
However, for these large deviations of the widths, we cannot expect good agreement with full numerical
simulations as is evident from the Fig.~\ref{fig_8}.
We note that the oscillations in the lowest panel in Fig.~\ref{fig_8} are unsymmetric w.r.t. $w_f(0)$ since the full spatial shapes (not presented here) of the clouds in the PDE simulations behaves quantitatively different at large amplitudes for fermions and bosons, as we have studied in detail numerically.
\begin{figure}
\includegraphics[scale=0.5]{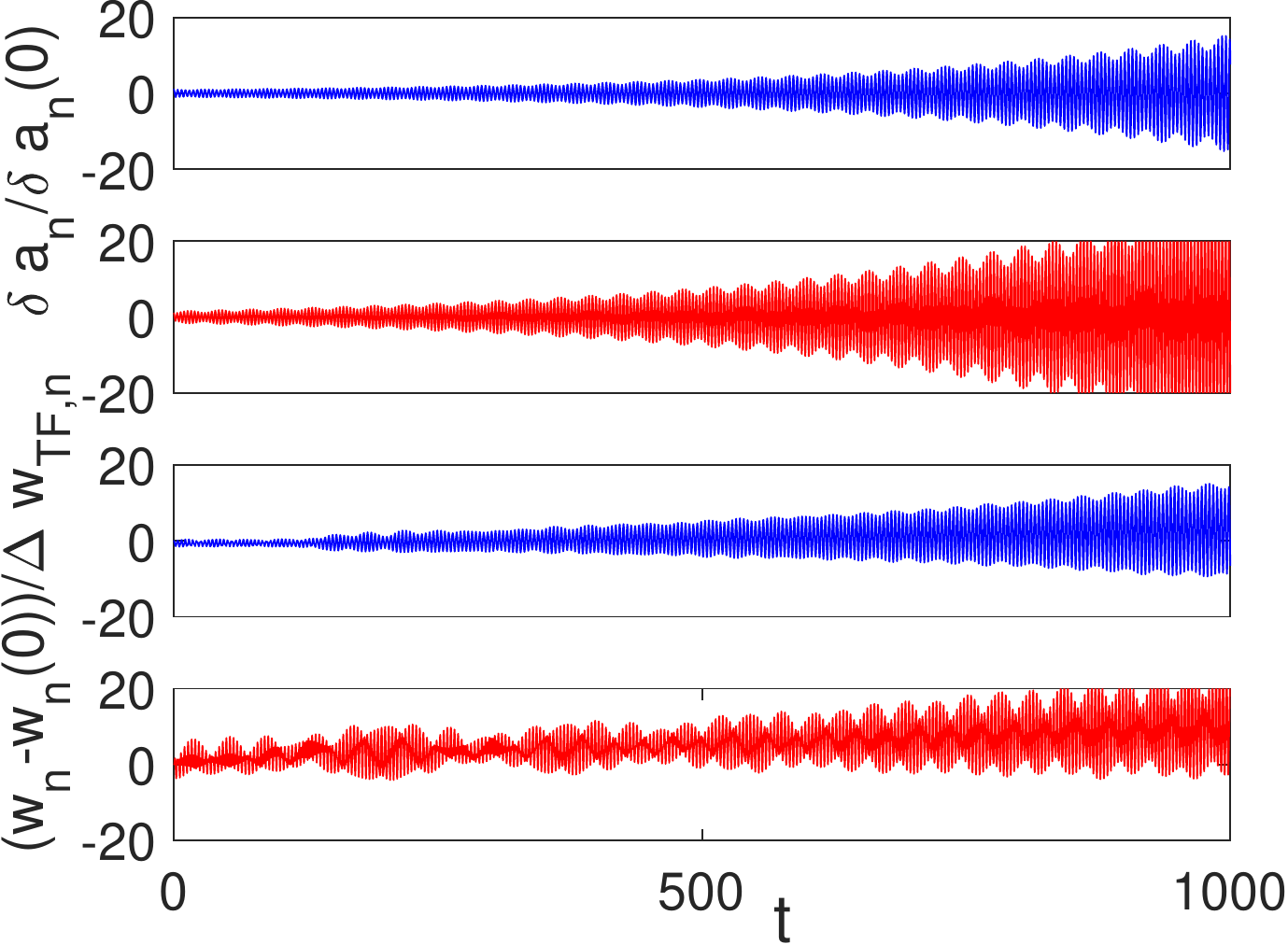}
\caption{(Color online) Parametric resonance of breathing modes, i.e. the change in the normalized RMS widths.
Two upper panels ($n=b$: blue; $n=f$: red): Solution to the ODE model, from Eqs.~(\ref{Eq_lin_a_b_a_f}) and~(\ref{BM_driving_frequency}), with parameters $g_{fb}^{(0)}=1/2$, $c_{fb}=0.2$, i.e. $\Omega_{fb}=\omega_1+\omega_2=3.6222$, and the initial values $\delta a_{b,f}(0)=0.005$.
The widths $a_f=a_{f0}$ and $a_b=a_{b0}$ are obtained numerically from Eq.~(\ref{equilibrium_widths}).
Two lower panels: Solution to the PDE model~(\ref{sys1}) with sinusoidal interspecies coupling~(\ref{TDgbANDg12}), with parameters $g_{fb}^{(0)}=1/2$, $c_{fb}=0.2$, $\Omega_{fb}=3.6358$.
An initial increase in the trapping frequences for both components was used (see caption of Fig.~\ref{fig_7}), such that the RMS widths initially have a change in the order $\Delta w_{TF,n} \simeq 0.005 w_{TF,n}$~\cite{Thomas-Fermi}.
Physical parameters are discussed in Sec.~\ref{Sec_Numerical_simulations}.
\label{fig_8}}
\end{figure}

\begin{figure}
\includegraphics[scale=0.47]{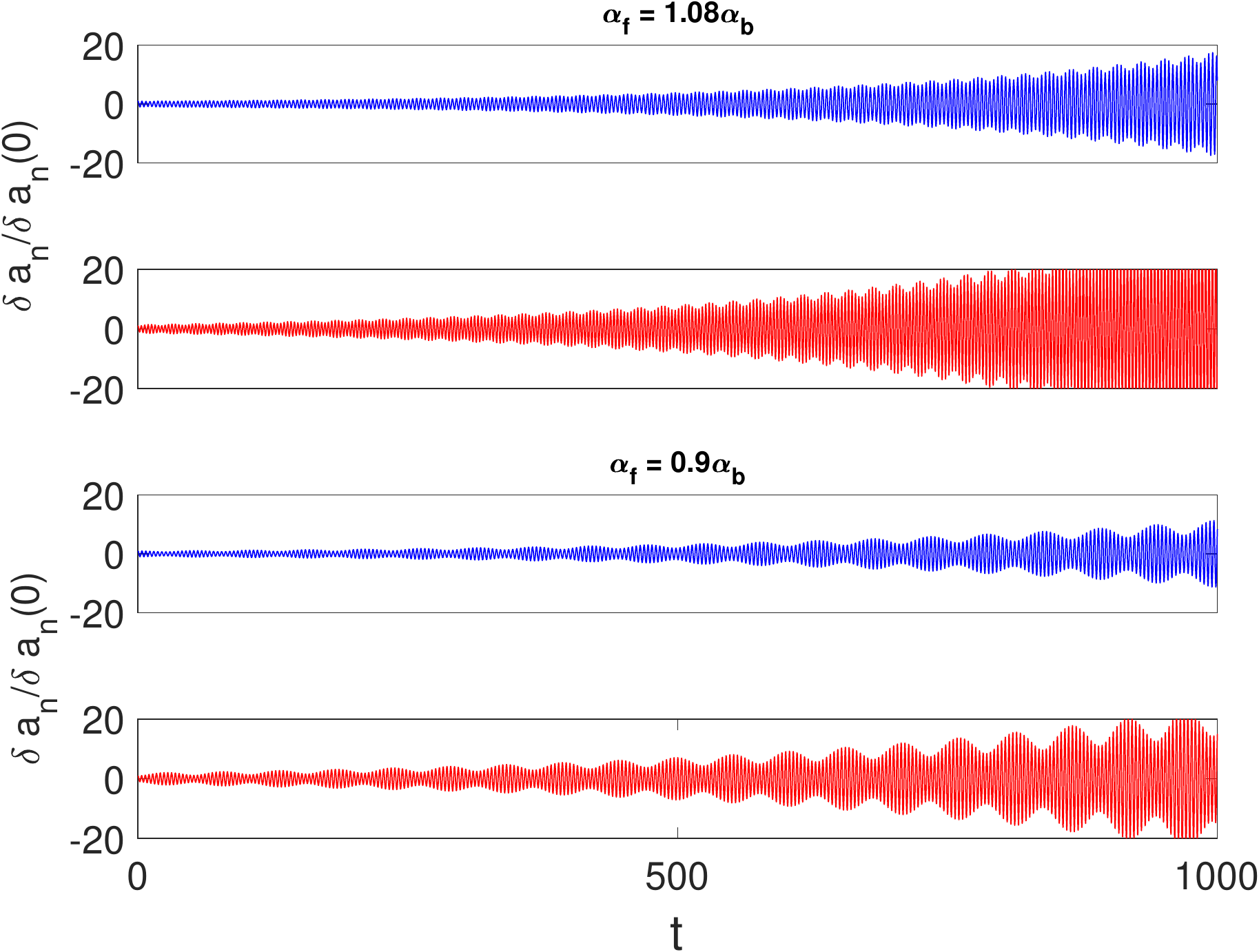}
\caption{(Color online) Parametric resonance of breathing modes for different traps $\alpha_b \neq \alpha_f$ within the ODE model.
Two upper panels ($n=b$: blue; $n=f$: red): Solution to the ODE model for $\alpha_b=1/4$ and $\alpha_f=1.08\alpha_b$, from Eqs.~(\ref{Eq_lin_a_b_a_f}) and~(\ref{BM_driving_frequency}), with parameters $g_{fb}^{(0)}=1/2$, $c_{fb}=0.2$, i.e. $\Omega_{fb}=\omega_1+\omega_2=3.6980$, and the initial values $\delta a_{b,f}(0)=0.005$.
The widths $a_f=a_{f0}$ and $a_b=a_{b0}$ are obtained numerically from Eq.~(\ref{equilibrium_widths}).
Two lower panels ($n=b$: blue; $n=f$: red): Solution to the ODE model according to the same procedure as for the upper panels but with a weaker trap for the fermions $\alpha_f=0.9\alpha_b$, i.e. $\Omega_{fb}=\omega_1+\omega_2=3.5232$.
Physical parameters are discussed in Sec.~\ref{Sec_Numerical_simulations}.
\label{fig_9}}
\end{figure}

\section{Numerical simulations} \label{Sec_Numerical_simulations}

Possible experimental parameters used for the calculations presented in the figures are the following:
we have illustrated the BCS regime ($\kappa=1/4$) for $N_b = 10^3$ number of bosons and $N_f  = 2 \cdot 10^2$ fermions, trapped with the same strength $\alpha_b=\alpha_f=1/4$ (except for in Figs.~\ref{fig_4},~\ref{fig_5} and~\ref{fig_9});
the intracoupling parameter in use was $g_b=g_b^{(0)}=1$ (see Sec.~\ref{sec_The_model} for the translation to dimensionless variables).

Furthermore, phase separation for the components is here expected for a coupling in the order of $g_{fb}\sim 1$~\cite{ViveritPRA2000}, which we have confirmed numerically.
Therefore, the curves in Figs.~\ref{fig_2} and~\ref{fig_7} based on the ODE model are dashed from this point on, and no rings from full PDE simulations are plotted, since we are not discussing the effect of phase separation here.

ODEs have been solved with Matlab's built in solver ode45 while PDEs have been solved with XMDS~\cite{XMDS} on dense enough $x-t$-grids.
The groundstates used for the initial conditions of the PDE~(\ref{sys1}) was calculated using a damped second order equation (DFPM)~\cite{SandinPRE2016}, using dynamical constraints~\cite{Gulliksson2018} to keep the number of particles constant.
Frequences of the oscillating observables have been obtained using exponential fitting as described in~\cite{AnderssonIEEE2014}, which is a development of the well known ESPRIT~\cite{roy1989esprit}.
In comparison with experimental data, additional noisereduction may be necesarry in obtaining the frequencies~\cite{andersson2017fixed}.

\section{Conclusion}
In conclusion we have investigated the collective oscillations of superfluid mixtures of ultra cold fermionic and bosonic atoms while varying the scattering lengths periodically in time.
The case of varying the fermion-boson $a_{s,fb}$ scattering lengths are studied with respect to excited center of mass modes and breathing modes in the mixture.
Parametric resonances in the oscillations are predicted and the properties are analyzed by comparing PDE and ODE models for the dynamics.
The resulting oscillations with increasing amplitudes provides clear experimental signals to search for.
A specific application is to the recent experiment with fermionic $^{6}$Li and bosonic $^{7}$Li atoms with oscillating fermion-boson scattering length, which can be realized using the Feshbach resonance technic.

\section{Acknowledgments}
F.~Kh.~A. acknowledges hospitality of \"{O}rebro University Sweden and partial support from a senior visitor fellowship from Conselho Nacional de Desenvolvimento Cientifico e Tecnol\'ogico (CNPq-Brasil).
The authors also acknowledge valuable comments from an anonymous referee.

\section{Appendix on the ODE model}
Variation with respect to each of the time dependent parameters $\xi_b$, $k_b$, $b_b$, $\xi_f$, $k_f$, and $b_f$ in the averaged Lagrangian (\ref{avlag1}) and invoking the Euler Lagrange equations
in (\ref{Euler-Lagrange-Equation}) results in the dynamical systems

\begin{eqnarray}
\label{A1}
\frac{d \zeta_b}{dt} &=& 2 k_b \nonumber \\
\frac{d a_b}{dt} &=& 4 a_b b_b \nonumber \\
\frac{d k_b}{dt} &=& -2 \alpha_b \zeta_b + 2 g_{fb} \frac{N_f}{\sqrt{\pi}} \frac{\zeta_b - \zeta_f}{(a_b^2 + a_f^2)^{3/2}} \exp \left( - \frac{(\zeta_b - \zeta_f)^2}{a_b^2 + a_f^2}\right) \nonumber \\
\frac{d \zeta_f}{dt} &=& 2 k_f \nonumber \\
\frac{d a_f}{dt} &=& 4 a_f b_f \nonumber \\
\frac{d k_f}{dt} &=& -2 \alpha_f \zeta_f + 2 g_{fb} \frac{N_b}{\sqrt{\pi}} \frac{\zeta_f - \zeta_b}{(a_b^2 + a_f^2)^{3/2}} \exp \left( - \frac{(\zeta_f - \zeta_b)^2}{a_b^2 + a_f^2}\right) \nonumber \\
\end{eqnarray}

\noindent These equations are coupled to the equations derived from variation with respect to
$A_b$, $a_b$, $A_f$ and $a_f$. The four final equations constitute a rather sizable system.
The first equation arises from variation with respect to $a_b$ and reads
\begin{eqnarray}
\label{A2}
&& \frac{3}{2} a_b^2 \frac{d b_b}{dt} + \frac{d \phi_b}{dt} = k_b^2 + \frac{1}{2 a_b^2} - 6 a_b^2 b_b^2 - \frac{3}{2} \alpha_b a_b^2 - \alpha_b \zeta_b^2 \nonumber \\
&& - \frac{g_b}{2 \sqrt{2}} \frac{N_b}{\sqrt{\pi} a_b} - g_{fb} \frac{N_f}{\sqrt{\pi}} (a_b^2 + a_f^2)^{-1/2} \exp \left( - \frac{(\zeta_b - \zeta_f)^2}{a_b^2 + a_f^2}\right) \nonumber \\
&& + g_{fb} a_b^2 \frac{N_f}{\sqrt{\pi}} \left( 1 - 2 \frac{(\zeta_b - \zeta_f)^2}{a_b^2 + a_f^2} \right) (a_b^2 + a_f^2)^{-3/2} \times \nonumber \\
&& \exp \left( - \frac{(\zeta_b - \zeta_f)^2}{a_b^2 + a_f^2}\right)
 = F_1(\xi_b, k_b, a_b, b_b, \xi_f, a_f) \; .
\end{eqnarray}

\noindent The second equation is derived by variation with respect to $a_f$ and reads
\begin{eqnarray}
\label{A3}
&& \frac{3}{2} a_f^2 \frac{d b_f}{dt} + \frac{d \phi_f}{dt} = k_f^2 + \frac{1}{2 a_f^2} - 6 a_f^2 b_f^2 - \frac{3}{2} \alpha_f a_f^2 - \alpha_f \zeta_f^2 \nonumber \\
&& - \frac{\kappa \pi}{3 \sqrt{3}} \frac{N_f^2}{a_f^2} - g_{fb} \frac{N_b}{\sqrt{\pi}} (a_b^2 + a_f^2)^{-1/2} \exp \left( - \frac{(\zeta_b - \zeta_f)^2}{a_b^2 + a_f^2}\right) \nonumber \\
&& + g_{fb} a_f^2 \frac{N_b}{\sqrt{\pi}} \left( 1 - 2 \frac{(\zeta_b - \zeta_f)^2}{a_b^2 + a_f^2} \right) (a_b^2 + a_f^2)^{-3/2} \times \nonumber \\
&& \exp \left( - \frac{(\zeta_b - \zeta_f)^2}{a_b^2 + a_f^2}\right)
 = F_2(\xi_f, k_f, a_f, b_f, \xi_b, a_b) \; .
\end{eqnarray}

\noindent The third equation is obtained from variation with respect to the Bose amplitude $A_b$
\begin{eqnarray}
\label{A4}
&& a_b^2 \frac{d b_b}{dt} + 2\frac{d \phi_b}{dt} = 2 k_b^2 - \frac{1}{a_b^2} - 4 a_b^2 b_b^2 - \alpha_b a_b^2 - 2\alpha_b \zeta_b^2 \nonumber \\
&& - \frac{g_b \sqrt{2} N_b}{\sqrt{\pi} a_b} - 2 g_{fb} \frac{N_f}{\sqrt{\pi}} (a_b^2 + a_f^2)^{-1/2} \exp \left( - \frac{(\zeta_b - \zeta_f)^2}{a_b^2 + a_f^2}\right) \nonumber \\
&& = F_3(\xi_b, k_b, a_b, b_b, \xi_f, a_f) \; .
\end{eqnarray}

\noindent The final and fourth equation results from variation with respect to the Fermi amplitude $A_f$
\begin{eqnarray}
\label{A5}
&& a_f^2 \frac{d b_f}{dt} + 2\frac{d \phi_f}{dt} = 2 k_f^2 - \frac{1}{a_f^2} - 4 a_f^2 b_f^2 - \alpha_f a_f^2 - 2\alpha_f \zeta_f^2 \nonumber \\
&& - \frac{2 \kappa \pi N_f^2}{\sqrt{3}a_f^2} - 2 g_{fb} \frac{N_b}{\sqrt{\pi}} (a_b^2 + a_f^2)^{-1/2} \exp \left( - \frac{(\zeta_b - \zeta_f)^2}{a_b^2 + a_f^2}\right) \nonumber \\
&& = F_4(\xi_f, k_f, a_f, b_f, \xi_b, a_b) \; .
\end{eqnarray}

\noindent For numerical implementation it is convenient to rewrite Eq. (\ref{A2}) and Eq. (\ref{A4}) into the form

\begin{eqnarray}
\label{A6}
\frac{d b_b}{dt} &=& \frac{1}{a_b^{2}} F_1 - \frac{1}{2 a_b^2}F_3 \; , \nonumber \\
\frac{d \phi_b}{dt}  &=& -\frac{1}{2} F_1 + \frac{3}{4} F_3 \; .
\end{eqnarray}

\noindent similarly for Eqs. (\ref{A3}) and \ref{A5})

\begin{eqnarray}
\label{A7}
\frac{d b_f}{dt} &=& \frac{1}{a_f^{2}} F_2 - \frac{1}{2 a_f^2}F_4 \; , \nonumber \\
\frac{d \phi_f}{dt}  &=& -\frac{1}{2} F_2 + \frac{3}{4} F_4 \; .
\end{eqnarray}

\noindent Note that these two final systems of equations are coupled to the system in (\ref{A1}).


\begin{thebibliography}{99}
\bibitem{Thomas-Fermi}
With $g_{fb}=0$ we have the standard Thomas-Fermi (TF) results for the atomic densities
$n_{TF,b}(x) = |\psi_b|^2 = \frac{1}{g_b} \left( \mu_b - \alpha_b x^2 \right)$ and
$n_{TF,f}(x) = |\psi_f|^2 = \sqrt{ \frac{1}{\kappa \pi^2} \left( \mu_f - \alpha_f x^2 \right) }$.
Using the TF radius $R_{TF,n} = \sqrt{\mu_n/\alpha_n}, \ n=f, \ b$ we can from the normalizations $N_n = \int n_{TF,n} dx$ obtain the standard TF widths
$R_{TF,b} =  \left( 3 g_b N_b /(4 \alpha_b) \right)^{1/3}$ and
$R_{TF,f} =  (\kappa/\alpha_f)^{1/4} \sqrt{ 2N_f}$,

while the corresponding TF RMS values that we compared to the ODE model are
$w_{TF,b}  = R_{TF,b}/\sqrt{5}$ and $w_{TF,f}  = R_{TF,f}/2$.
\end{thebibliography}
\end{document}